\documentclass[a4paper,fleqn,usenatbib]{mnras}
\usepackage{graphicx,amsmath,amssymb,caption}
\newcommand{\be} {\begin{equation}}
\newcommand{\ee} {\end{equation}}
\newcommand{\bea}{\begin{eqnarray}}
\newcommand{\eea}{\end{eqnarray}}
\def\aap{A\&A}
\def\apj{ApJ}

\def\apjs{ApJ S}
\def\mnras{MNRAS}

\def\icarus{Icarus}
\def\araa{ARAA}

\begin{document}
\title{Tidal Evolution of Close-in Exoplanets and Host Stars}
\author[S. Ferraz-Mello and C. Beaug\'e]{S. Ferraz-Mello$^{1}$\thanks{E-mail: 
sylvio@usp.br}, C. Beaug\'e$^{2}$ \\
$^{1}$ Instituto de Astronomia, Geof\'{\i}sica e Ci\^encias Atmosf\'ericas, 
Universidade de S\~ao Paulo, Brasil \\
$^{2}$ Instituto de Astronom\'{\i}a Te\'orica y Experimental, Observatorio 
Astron\'omico, Universidad Nacional de C\'ordoba, Argentina\\
}
\date{}
\pubyear{2023}
\label{firstpage}
\pagerange{\pageref{firstpage}--\pageref{lastpage}}
\maketitle

\begin{abstract}
The evolution of exoplanetary systems with a close-in planet is ruled by the 
tides mutually raised on the two bodies and by the magnetic braking of the host 
star. This paper deals with consequences of this evolution and some features  
that can be observed in the distribution of the systems two main periods: the 
orbital periods and the stars rotational periods. The results of the 
simulations are compared to plots showing both periods as determined from the 
light curves of a large number of Kepler objects of interest. These plots show 
important irregularities as a dearth of systems in some regions and 
accumulations of hot Jupiters in others. It is shown that the accumulation of 
short-period hot Jupiters around stars with rotation periods close to 25 days 
results from the evolution of the systems under the joint action of tides and 
braking, and requires a relaxation factor for solar-type stars of around $10 \, 
s^{-1}$.
\end{abstract}


\section{Introduction}\label{intro}

One difficulty faced by tidal evolution theories is the extremely slow time 
scale of these phenomena and the almost impossibility of observing them at work. 
So far, tidal infall rates have only been estimated for two exoplanets: WASP-12b 
\citep[$ 29 \pm  2$ ms/year cf.][]{Yee.etal.2020} and Kepler-1658b \citep[$ 131 
\pm  20$ ms/year cf.][]{Vissapragada.etal.2022}, both obtained from analysing 
variations in the orbital period of the planet. For all other systems, tidal 
evolution must be deduced indirectly from present-day dynamical structures that 
may have been generated or affected by tidal interactions.

Together with magnetic braking, the transfer of angular momentum from the 
companion orbit to the rotation of the host star plays an important role 
in defining the rotational and orbital evolution of the system
\citep[e.g.,][]{Bolmont.etal.2012,Teitler.Konigl.2014,Ferraz-Mello.etal.2015}. 
Key elements in this case are the physical parameters of the star and of the 
planetary companion, and the age of the system - one important datum of 
difficult determination. We may also use our knowledge of the distribution of 
some physical parameters to infer some properties of the tidal models. One 
example is the analysis done by \cite{Hansen.2010} using the distribution of 
eccentricities, periods and masses of the known planets, which allowed him to 
constrain the dissipation values of stars and planets.     

\begin{figure*}
\begin{center}
\resizebox{2.05\columnwidth}{!}{\includegraphics{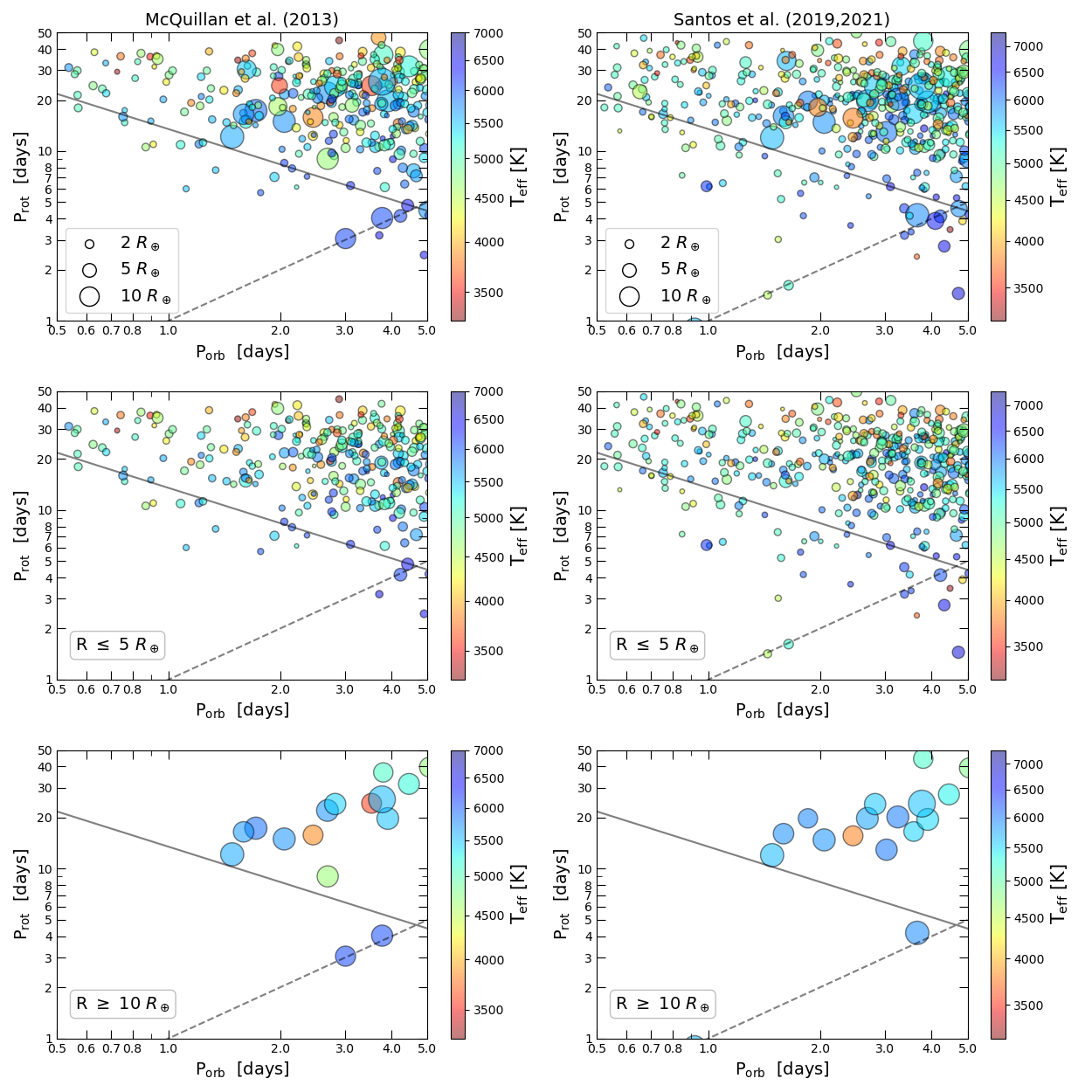}}
\caption{{\bf Left:} Period of the stellar rotation as a function of the 
orbital period for confirmed planets and KOIs with measured rotation periods 
and orbital period less than 5 days \citep[after][]{McQuillan.etal.2013}. In 
the middle and bottom frames the data is restricted to relatively small 
($R < 5 R_\oplus$) and large ($R > 10 R_\oplus$) planets, respectively. Dashed 
lines identify conditions for synchronous rotation ($P_{\rm orb} = P_{\rm rot}$) 
while continuous lines reproduce the lower envelope of points proposed by 
\protect\cite{McQuillan.etal.2013}. {\bf Right:} Same as the left-hand column, 
but using updated planetary and stellar data from 
\protect\cite{Santos.etal.2019,Santos.etal.2021}.}
\label{fig1}
\end{center}
\end{figure*}

The top left-hand frame of Figure \ref{fig1} is an excerpt of the full figure 
published by \cite{McQuillan.etal.2013} and based on the three first years of 
public Kepler data. It shows the distribution of stellar rotational periods 
$P_{\rm rot} = 2\pi/\Omega_0$ and planetary orbital periods $P_{\rm orb} = 
2\pi/n$; hereafter the P-P diagram. As shown in the inset, the size of 
each circle is proportional to the planetary radius, while its color is 
determined by the effective temperature of the host star. We note an almost 
absence of systems with fast rotating host stars and $T_{\rm eff} < 5800$ K. 
This last characteristic is just due to the loss of angular momentum of the 
colder stars due to stellar winds, the so-called magnetic braking 
\citep[see][]{Bouvier.etal.1997}. The hottest stars do not have the outer 
convective layer responsible for the winds and so do not migrate upwards in this 
figure as the colder stars. 

As indicated by \cite{McQuillan.etal.2013}, the large majority of close-in 
planets ($P_{orb} < 5$ days) are located above the continuous line drawn in all 
panels of Figure \ref{fig1}, while a few also appear to lie close to a 
synchronous state ($P_{orb} = P_{rot}$) highlighted with a dashed line. The 
region between both these lines appears surprisingly empty.

The same feature is found when restricting the data to smaller planets, as seen 
in the middle left-hand frame, where only bodies with $R < 5 R_\oplus$ are 
shown. Several explanations were proposed for the lack of planets below the 
continuous line, including the combined effects of tidal evolution, braking and 
magnetic star-planet interactions \citep[e.g.,][]{Teitler.Konigl.2014, 
Ahuir.etal.2021} as well as dynamical analysis of tidal capture of 
near-parabolic bodies in multi-planet systems \citep{Lanza.Shkolnik.2014}. None 
of these works, however, focus on the distribution of large planets, seen here 
in the lower left-hand plot of Figure \ref{fig1}. Not counting the two KOIs 
close to the synchronous line, most bodies seem to show a significant 
correlation, with larger stellar rotation associated to larger orbital periods 
and lower values of $P_{rot}$ for hot Jupiters closer to the star.

\begin{table*}
\centering
\caption{Hot exoplanets -- Adopted parameters} 
\begin{tabular}[h!]{lccccccc}
\hline
Planets &  Mass & \multicolumn{2}{c} {Radius} & $\alpha$ & $k_f$ & $\gamma$ 
\\
& $(m_{\rm Jup})$& $(R_{\rm Jup})$ &$(R_{\rm Earth})$& $(mR^2)$ & & (${\rm 
s}^{-1}$)\\
\hline
Jupiters       & 1       & 1      & 11.2 & 0.254 & 0.38 & 20 \\
Saturns        & 0.3     & 0.843  & 9.5  & 0.21  & 0.34 & 20 \\
mini-Saturns   & 0.15    & 0.54   & 4.7  & 0.21  & 0.34 & 20 \\
Neptunes       & 0.0598  & 0.346  & 3.9  & 0.23  & 0.12 & 10 \\
mini-Neptunes  & 0.02    & 0.24   & 2.7  & 0.23  & 0.12 &  5  \\
super-Earths   & 0.0126  & 0.1424 & 1.6  & 0.33  & 0.3  &  5$\times 10^{-7}$\\
Earths         & 0.00315 & 0.089  & 1    & 0.33  & 0.3  &  5$\times 10^{-7}$\\
\hline
\end{tabular}\label{tab:data}
\end{table*}

The right-hand frames of Figure \ref{fig1} show the same distributions in 
the P-P diagram, but employing more recent data. Stellar rotations for main 
sequence M and K stars are taken from \cite{Santos.etal.2019} while similar 
information for G and F stars are found in \cite{Santos.etal.2021}. These were 
then compared with the latest catalogue of Kepler candidates, while the same 
database also allowed to identify and remove possible eclipsing binaries. 
The total number of data points increased from 1079 to 1698. Even so, it is 
important to keep in mind that many correspond to Kepler candidates (KOI) and 
therefore do not constitute confirmed planets. This explains why some data 
points in the left-hand column are absent in the latest catalogues.

While some characteristics of the distribution in the P-P diagram remain, others 
have undergone noticeable changes. The diagonal continuous line now appears less 
decisive as a lower bound for orbital periods of small close-in planets, and the 
middle right-hand plot shows several new systems for a wider of range of stellar 
rotations. In fact, more recent data from both Kepler and TESS 
\citep{Messias.etal.2022} suggest that the dearth of close-in small planets 
around rapidly rotating stars could be due to a lack of data and thus not 
statistically significant. 

Conversely, the correlation between $P_{\rm orb}$ and $P_{\rm rot}$ in hot 
Jupiters (lower right-hand plot) appears even more pronounced and better 
defined, and practically all bodies are located in a moraine-like accumulation. 
The only exception is KOI 554, found very close to the synchronous line. This 
system appears as unconfirmed in NASA's lists and is a probable false positive.

In this paper we present a simple dynamical model to analyze the orbital and 
rotational evolution of a single planet around a star, under the combined 
effects of tidal interactions and magnetic braking. We will show that the 
observed distribution of small planets as well as the accumulation observed for 
giant planets can be well reproduced, and, at least for Solar-type stars, allow
for a reliable estimation for the stellar relaxation factor $\gamma$.

\section{The Dynamical Model}\label{sec:Bj}

We computed the evolution tracks of close-in planets taking into account tidal 
effects and the magnetic braking of the star. Braking of the stellar rotation 
was computed following \citep{Bouvier.etal.1997} while the tidal evolution of 
both the planet and host star were calculated using the creep tide theory 
\citep{Ferraz-Mello.2013,Ferraz-Mello.2015}. Equations and details of the 
dynamical model are given in Appendix \ref{sec:eqs}. 

The codes adopted for this paper differ slightly from those used in the study of 
the interplay of tidal evolution and stellar wind braking in the rotation of 
stars hosting massive close-in companions \citep{Ferraz-Mello.etal.2015}. The 
main improvements include the introduction of the actual fluid Love 
number $k_f$ so that the equations correspond to the case of a differentiated 
body whose layers are aligned homogeneous ellipsoidal shells and the 
introduction of the effects due to the shortening of the polar axis by the tidal 
potential \citep{Ferraz-Mello.2015}.

\begin{figure*}
\begin{center}
\resizebox{2.05\columnwidth}{!}{\includegraphics{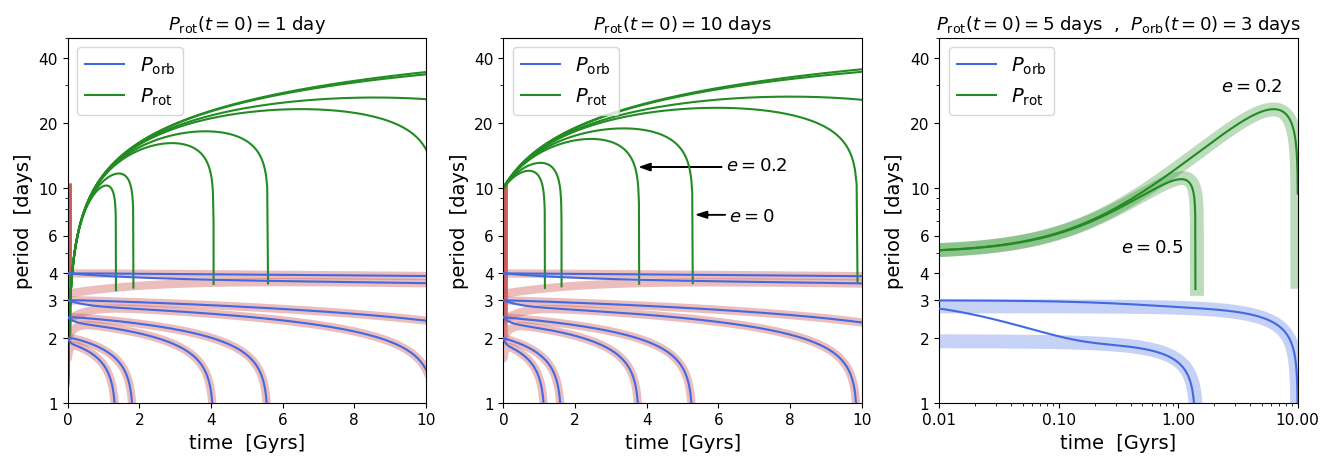 } }
\caption{Dynamical evolution of a Jupiter-size planet orbiting a Solar-type 
star under tidal interactions and magnetic braking. Green lines show the 
evolution of the stellar rotational period $P_{\rm rot}$ while blue lines show 
the planet's orbital period $P_{\rm orb}$ . Broad pink curves in the left and 
center plots follow the change in the planet's rotational period. {\bf Left \& 
Center:} Initial value of $P_{\rm rot}$ fixed to $1$ and $10$ days, 
respectively, while four initial planet orbital periods were analyzed ($P_{\rm 
orb} = 2,2.5,3,4$ days). In each case we considered two initial values of $e$, 
identified in the center plot. {\bf Right:} Thin continuous lines show the time 
evolution of both $P_{\rm rot}$ and $P_{\rm orb}$ for the same initial values 
(specified on top) but two different planetary eccentricities. Broad light 
colored lines show the evolution of circular synchronized planets 
($P_{\rm orb} = 2\pi/n$) with initial orbital periods chosen to reproduce 
the orbital decay after circularization. See text for details.}
\label{fig2}
\end{center}
\end{figure*}

In all the simulations, the physical set-up was assumed frozen. Certainly, some
parameters involved in the calculations are expected to vary during the whole 
simulated story. The mass of the star is expected to vary a small amount during 
the system lifetime. This variation could be included in the simulations, but 
would not be able to change significantly the results. Large variations may be 
expected in the planetary masses if the planet is too close to the star because 
of evaporation of its outer layers. However, the influence of these variations 
in the simulation results is negligible. The variation in the orbital parameters 
when masses are not constant is proportional to the derivative of the sum of the 
two masses \citep{Hadjidemetriou.1963}; the planet has nothing but a tiny 
fraction of the total mass of the system and so, the sum of the masses will be 
almost unaffected by variations in the mass of the planet even if they are 
large. 

Other parameters showing variation with the evolution of the star are the fluid 
Love number and the radius of the star. They are related to the density profile 
of the star. The radius of the star may have a significant variation, but it 
appears in the equations only through $k_f R^2$; Models of the evolution of the 
internal structure of the stars \citep{Claret.2019} show that the product $k_2 
R^2$ and the moment of inertia have just a small variation during the time in 
which the stars remains in the main sequence and variations can be neglected.
More complex models taking into account the variations in the internal structure 
of the star would introduce new non-universal unknown parameters without 
introducing significant changes in the results.

Table \ref{tab:data} summarizes the main physical parameters adopted for 
each type of planet, including nominal values for the relaxation factor 
$\gamma$. We denote by $\alpha$ the multiplicative factor in the expression 
of the body's moment of inertia, i.e. $C = \alpha mR^2$. The adopted values of 
the planetary $k_f$ are based on those calculated for similar Solar System 
planets by \cite{Gavrilov.Zharkov.1977}. Different choices for $k_f$ affect the 
estimations of the relaxation factor $\gamma$ since these two quantities are 
entangled in the tidal model. In a first approximation, for gaseous bodies, the 
tidal variation of the elements is proportional to the ratio $k_f/\gamma$ (see 
Appendix \ref{sec:eqs}). However, in the present case, the contribution of the 
tides on the planet to the variation of the orbital period is insignificant and 
the planetary $\gamma$ and $k_f$ are given only to complete the information on 
the parameters used in the simulations.

Figure \ref{fig2} shows the results of some preliminary simulations involving a 
Jupiter-like planet orbiting a Solar-type star. While the planetary parameters 
were taken from Table \ref{tab:data}, for the central mass we adopted a moment 
of inertia, $C_0=0.07 m_0 R_0^2$, similar to the present-day Sun. Its fluid Love 
number was chosen equal to $k_{f_0}=0.05$, as obtained using the equations 
derived by \cite{Folonier.etal.2015}, and assuming a density profile equal to 
the standard solar model. The stellar relaxation factor was fixed at $\gamma=10$ 
$s^{-1}$, in the middle of the range indicated by previous studies 
\citep[see][]{Ferraz-Mello.2022}, but corrected for the smaller value of 
$k_{f_0}$. 

The left-hand and center plots show, in blue, the time evolution of the planet's 
orbital period $P_{\rm orb}$ for four different initial values: $2 \, , \, 2.5 
\, , \, 3$ and $4$ days. For each case we considered two different initial 
eccentricities: $e = 0$ and $e = 0.2$, while the initial rotation frequency of 
the planet was taken equal to $2\pi/\Omega_1 = 10$ days. This value is 
arbitrary, but we found no significant difference as long as the body was 
initially sub-synchronous ($P_{\rm orb} < P_{\rm rot}$). Finally, both graphs 
differ in the initial rotational period of the star, as indicated by the text on 
top. Fast rotators are considered on the left while initial slow rotators are 
considered for the center plot. The tidal evolution was followed according to 
the equations described in the Appendix, considering both the star and planet as 
extended bodies and including Cayley functions $E_{0,k}$ and $E_{2,k}$ up to 
order $k = 7$.

The pink lines show the evolution of the planetary rotational period. The 
orbital circularization, together with the synchronization of the planetary spin 
($\Omega_1 = n$), occurs early in the evolution of the system, as indicated by 
the superposition of the pink and blue lines. The longest timescale is found for 
$P_{\rm orb}(t=0) = 4$ days, where the synchronization occurs after $1$ Gyr. 
However, the subsequent orbital evolution is almost negligible even after $T = 
10$ Gyrs. The stellar rotation period, however, fueled by magnetic braking, 
increases monotonically, reaching values of the order of $30$-$40$ days at the 
end of the simulation. 

Planets closer to the star tell a different story. Synchronization occurs very 
early in the system's history and most of the changes in $P_{\rm orb}$ occur in 
a scenario dominated by stellar tides and where the planetary counterparts are 
negligible. Tidal evolution for these initial conditions are much stronger than 
those for $P_{\rm orb}(t=0) = 4$ days, and the bodies are engulfed by the star 
in timescales between $1$-$10$ Gyr. Conservation of angular momentum implies 
that the orbital infall of the planet also causes a decrease in the rotational 
period of the star. Given the large planet mass adopted for these runs, this 
effect is able to counteract the magnetic braking and at some point during the 
system's evolution $P_{\rm rot}$ peaks and starts to decrease in value. Although 
this behavior is well known \citep[e.g.][]{Ferraz-Mello.etal.2015}, its effect 
on the distribution of hot Jupiters in the P-P plane has yet to be explored. 

A comparison between the left and center plots seem to indicate that the 
evolution of the system is only weakly dependent on the star's initial 
rotational period. The orbital evolution of the planet appears virtually 
identical in both cases, except for a change in the timescale of the order of 
$0.5$ Gyr, The values of $P_{\rm rot}$ are of course different, but both tend 
to very similar values after $\sim 5$ Gyrs. However, a more complex dependence 
is noted with respect to the initial eccentricity. 

The thin continuous lines in the right-hand side plot of Figure \ref{fig2} shows 
the results of two simulations with the same initial values for $P_{\rm orb}$ 
and $P_{\rm rot}$ (indicated on top) but different initial eccentricities. While 
the general trend is similar in all cases, the timescale and the maximum value 
attained by $P_{\rm rot}$ varies significantly, even though the planet reaches a 
synchronous state with almost circular orbit before $0.2$ Gyrs. This result is 
interesting since it indicates very different outcomes and timescales even 
though most of the system's evolution occurs in circular orbit and dominated by 
stellar tides. 

An explanation may be found precisely during the road towards synchronization. 
Since the planet begins with $P_{\rm orb} < P_{\rm rot}$, the early orbital 
decay rate is much stronger than that expected for $\Omega_1 \sim n$. This may 
be observed in the behavior of $P_{\rm orb}$ during the first stage of the 
system's evolution. Once synchronization is reached, the orbital decay levels 
out and the subsequent decay is much shallower. The broad light-colored lines 
show the evolution of ``equivalent'' systems, characterized by $e = 0$, 
$\Omega_1 = n$ and initial semimajor axis $a_{\rm equiv}$ chosen such 
that both the planet's orbital and star's rotational evolution mimics the 
original system after synchronization. For initial rotation rates such that 
$\Omega_1 \ll n$ we found $a_{\rm equiv} \simeq a_{\rm ini} \left( 1 - e_{\rm 
ini}^2 \right)$, compatible with the conservation of the orbital angular 
momentum. 

The overlap between the evolution of eccentric systems and their equivalent 
counterparts has far-reaching consequences. The most obvious is that we can 
simulate the tidal/braking interaction assuming circular orbits and 
synchronized planets as long as we adopt $a_{\rm equiv}$ as the initial 
semimajor axis. The tidal equations are thus simpler, the number of differential 
equations are reduced and the numerical integration run much faster. 

The second consequence is more relevant to our study and, perhaps, more 
debatable. Basically, we may say that when analyzing the origin of the observed 
distribution of exoplanets in the P-P diagram, we need not be concerned about 
the primordial distribution of semimajor axes and eccentricities, but solely 
the distribution of $a_{\rm equiv}$. We can thus approach our problem using 
the simplified tidal equations as long as we keep in mind that the initial 
separation between star and planet must be considered as representative and not 
equal to the true primordial value.

\section{Evolution of systems with a close-in gaseous giant}\label{sec:Bj}

We can now proceed to analyze whether our simple dynamical model can explain the 
distribution of planets in the P-P diagram, and see what tidal parameters are 
most suited for such a process. We begin studying hot Jupiters around Solar-type 
stars. As shown in Figure \ref{fig1}, the distribution of large ($R > 10 
R_\oplus$) KOIs taken from \cite{Santos.etal.2019,Santos.etal.2021}, defines a 
moraine-like accumulation with a positive correlation between $P_{\rm orb}$ and 
$P_{\rm rot}$, at least for close-in planets (or planetary candidates) with 
orbital periods up to $5$ days. The only exception is KOI 554 which, as 
discussed previously, is probably a false positive. The same overall 
distribution is also found in older data, such as \cite{McQuillan.etal.2013} 
although perhaps less streamlined. 

There are currently four different mechanisms proposed to explain the origin of 
hot Jupiters. Tidal capture following eccentricity excitation from 
planet-planet scattering \citep{Beauge.Nesvorny.2012} or perturbations from a 
stellar companion \citep{Naoz.etal.2012}, disc-induced planetary migration 
\citep{Crida.Batygin.2014} and secular chaos \citep{Wu.Lithwick.2011}. The 
distribution of misalignment angles and planet multiplicity seem to indicate 
that no single process acted alone and at least part of the known population 
is a consequence of disc-planet interactions while the rest may be traced to 
orbital circularization after tidal capture \citep{Dawson.Johnson.2018}. 
Among the different dynamical processes leading to tidal capture, 
the distribution of misalignment angles seems to favor planet-planet scattering 
\citep{Marti.Beauge.2015} which may have occurred shortly after the dissipation 
of the gaseous disk \citep{Lega.etal.2013,Izidoro.etal.2021}. 

Both disk-planet interactions and post-dissipation instabilities point towards 
an early accumulation of hot Jupiters in their current orbital distance and, 
consequently, we can assume that the central star was still a rapid rotator at 
that time. Although the slow process of secular chaos may not be ruled out, it 
is currently difficult to evaluate how much it may have contributed to the 
observed distribution. Thus, even though we cannot rule out that some hot 
Jupiters may have reached their current orbital distance after magnetic braking 
drove the star to a slow rotation rate, there is little evidence to indicate 
that their number was substantial. 

\begin{figure}
\begin{center}
\resizebox{\columnwidth}{!}{\includegraphics{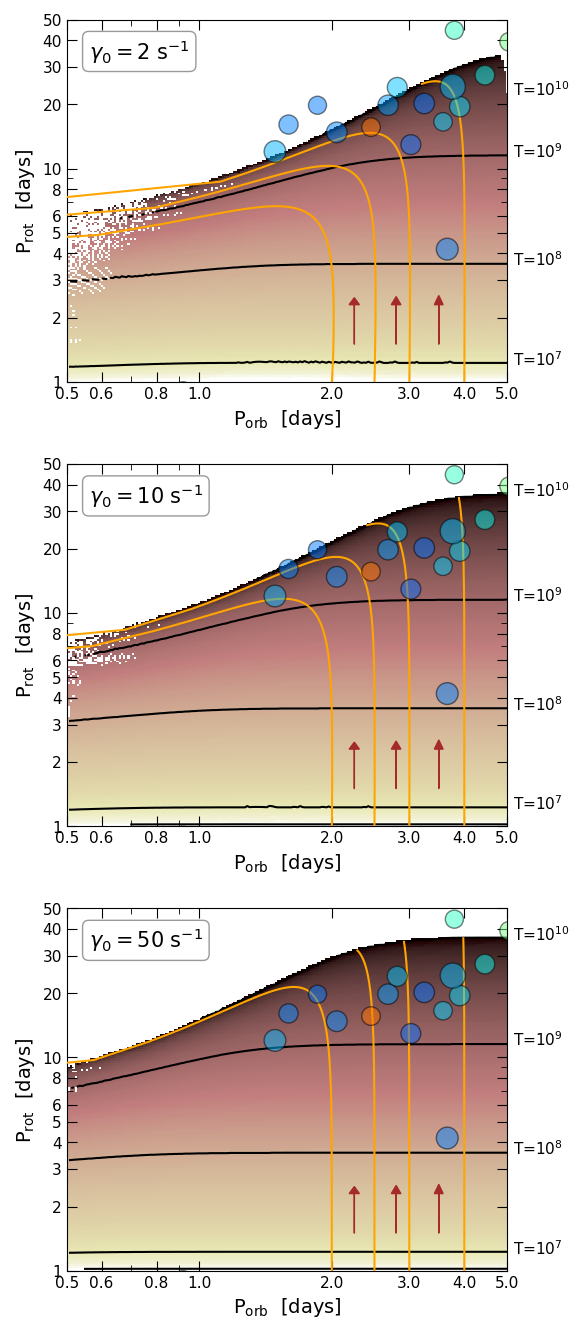}}
\caption{Overlaid to the distribution of hot Jupiters in the P-P diagram, each 
plot shows the evolutionary tracks of $4000$ initial conditions for 
Jupiter-size planets with $e = 0$, $\Omega_1=n$ and orbital periods in the 
interval $P_{\rm orb} = 2\pi/n \in [0.5,5]$ days, We assume a Solar-type star 
with initial rotation period $P_{\rm rot} = 1$ day. Each frame considers a 
different stellar relaxation factor $\gamma_0$ highlighted in the top left-hand 
corner. The color code is indicative of the age of the system $T$ (in years) as 
it transverses the plane following the arrows. Orange curves show the evolution 
of four representative initial conditions.}
\label{fig3}
\end{center}
\end{figure}

The three frames of Figure \ref{fig3} show, for different values of the stellar 
relaxation factor $\gamma_0$, the evolution of $4000$ initial conditions in the 
P-P diagram, all consisting of a Jupiter-size planet in circular orbit and 
initial orbital period in the range $P_{\rm orb} \in [0.5,5]$ days. The central 
star was assumed Solar-type with with an initial stellar rotation $P_{\rm rot} = 
1$ day. The orange lines show the evolutionary tracks of four characteristic 
initial conditions; their starting orbital periods were chosen to be $2$, $2.5$, 
$3$ and $4$ days, respectively. Since we are assuming initial circular orbits 
and synchronous planets, the initial values of $P_{\rm orb}$ correspond to the 
equivalent semimajor axes, as defined in the previous section.

The arrows indicate the flow of the system. The characteristic evolution
timescale for stellar rotation, driven by magnetic braking, is much shorter
than the evolution timescale for the orbital period of the planet that is
driven by tides. Therefore, even for an initial stellar rotation period shorter
than the initial orbital periods, the evolutionary tracks begin as almost
vertical straight lines. This stage, however, is temporary and after 
$10^8-10^9$ years the system reaches the domain of the accumulation highlighted 
in Figure \ref{fig1}. For very small semimajor axes the rotation of the star is 
then accelerated due to the transfer of orbital angular momentum to the rotation 
of the star by means of the tides raised by the planet on the star. The 
planet falls on the star in a few Gyrs. 

For wider systems, the tidal effects are not strong enough to cause the infall 
of the planet, but the evolution almost stops after $\sim 6$ Gyr and the 
systems no longer shows a significant evolution in the P-P diagram. These hot 
Jupiters accumulate in a moraine-like structure. Like in the geological process 
with this name, big planets are carried along with a flow to the domain where 
the flow becomes weaker and accumulate there. 

Since the evolutionary tracks of different initial conditions do not cross, we 
may color-code the P-P diagram indicating the age at which the system reached 
a given spot. Lighter tones correspond to early stages in the evolution, while 
the dark brown region highlights the position attained by all initial conditions 
at times between $1$ and $10$ Gyrs, the estimated range of ages for these 
systems \citep{Lanza.Shkolnik.2014}. To further aid in following the time 
evolution of the systems, the black lines show the isochrones for 
$\log_{10}(T) = 7,8,9,10$, where the time is given in years.

These results show that our simple dynamical model (tidal evolution + magnetic 
braking) lead to a distribution of Jupiter-size planets in the P-P plane that 
strongly resembles the observed distribution of hot Jupiters. The moraine shape 
is a natural consequence of the interplay between both phenomena on the stellar 
rotation, and may be used to estimate the value of the stellar relaxation 
factor $\gamma_0$ that appears to lead to a better correlation. As shown in the 
top frame of Figure \ref{fig3}, a value of $\gamma_0 = 2$ $s^{-1}$ generates an 
excessively efficient tidal decay and a significant portion of the observed hot 
Jupiters lie above the isochrone associated to $\log_{10}(T) = 10$. The larger 
relaxation factor considered in the middle frame leads to a much better fit, 
with practically all the population embedded in the region between $T=10^9$ 
yrs and $T=10^{10}$ yrs. A similar conclusion may be drawn from the lower plot 
where an even larger value of $\gamma_0$ is considered. 

We may thus deduce from these graphs that the observed distribution of hot 
Jupiters is consistent with a tidal evolution dominated by stellar tides 
with a relaxation factor between $10$ and $50$ $s^{-1}$. As the orbital distance 
of the planet decreases, so does the stellar rotation, leading to a 
moraine-type accumulation in the P-P diagram. Consequently, for massive close-in 
planets the stellar rotational period does not necessarily grow monotonically 
with time and care must be taken when using $P_{\rm rot}$ as a direct proxy for 
stellar age. 

As we mentioned at the beginning of this analysis, most of the hot Jupiters are 
expected to have reached their present location before the stellar rotation 
slowed significantly. Even if this was not the case, we have found that the
initial rotational period of the star exerts almost no influence on the 
evolution. As shown in the simulations carried out in Figure \ref{fig2}, even a 
relatively slow initial rotation of $10$ days led to the same evolutionary 
tracks in the P-P plane after $\sim 1$ Gyrs. Even the final system ages for a 
given value of $P_{\rm orb}$ or $P_{\rm rot}$ only varied by $\sim 0.5$ Gyr at 
the end of the simulations. We therefore believe that the above analysis should 
be fairly robust and independent of the initial conditions.

\section{Evolution of systems with a close-in Sub-Jovian planet}\label{sec:Bu}

Regardless of which stellar relaxation factor better fits the data, it seems 
that the moraine is (at least partially) caused by a funneling of the 
dark-toned region as $P_{\rm orb} \rightarrow 0$. In turn, this effect is 
associated to a change in sign of the time derivative of the star rotation 
$P_{\rm rot}$. Regardless of the age of the system when this occurs, as soon as 
the tidally-induced speed-up of the stellar rotation surpasses the slow-down 
caused by magnetic braking, the evolutionary tracks converge and lead to an 
orbital infall of the planets in a tight formation. 

To understand when this phenomena occurs and under what system parameters, we 
look for solutions of the equation:
\be
\left. \frac{d\Omega_0}{dt} \right|_{\rm tid} + \left. \frac{d\Omega_0}{dt} 
\right|_{\rm mag} = 0 ,
\label{eq2}
\ee
where the first term is the derivative of the stellar spin $\Omega_0$ due to 
stellar tides and the second is the contribution from magnetic braking. Full 
expressions for both are given in Appendix \ref{sec:eqs}. Assuming 
$|\nu|/\gamma_0 = 2|n-\Omega_0|/\gamma_0 \ll 1$ and $n \gg \Omega_0$, we 
can approximate the tidal term by:
\be
\left. \frac{d\Omega_0}{dt} \right|_{\rm tid} \simeq \frac{3}{\gamma_0} 
\frac{k_{f_0}}{\alpha_0} \left( \frac{m_1}{m_0} \right)^2 
\left( \frac{R_0}{a} \right)^3 n^3 .
\label{eq3}
\ee
Similarly, since we expect the moraine to occur for slow rotators where the 
magnetic braking is not so efficient, for this region we can approximate the 
second term by:
\be
\left. \frac{d\Omega_0}{dt} \right|_{\rm mag} = -B_w \Omega_0^3.
\label{eq4}
\ee
Introducing both expressions into (\ref{eq2}) we obtain that the maximum 
stellar rotational period $P^{(\rm max)}_{\rm rot}$, before tidal effects begin 
to dominate, is approximately given by:
\be
\left( P^{(\rm max)}_{\rm rot} \right)^3 \simeq \frac{B_w \gamma_0}{3}
\frac{\alpha_0}{k_{f_0}} \left( \frac{m_0}{m_1} \right)^2 
\left( \frac{a}{R_0} \right)^3 P_{\rm orb}^3 .
\label{eq5}
\ee
Relating $a$ and $P_{\rm orb}$ through Kepler's third law, and taking base 
$10$ log on both sides, we finally obtain
\be
\log_{10}{P^{(\rm max)}_{\rm rot}} \simeq \frac{5}{3} \log_{10}{P_{\rm orb}} + 
\frac{1}{3} \log_{10}{\Lambda}, 
\label{eq6}
\ee
where $\Lambda$ is a constant that depends only on the physical properties of 
the system, and is approximately given by
\be
\Lambda \simeq \frac{B_w \gamma_0}{3} \frac{\alpha_0}{k_{f_0}} \frac{G 
m_0}{R_0} \left( \frac{m_0}{m_1} \right)^2 . 
\label{eq3}
\ee
We thus obtain a linear relation between both periods in a log-log scale; 
moreover for any given value of $P_{\rm orb}$ the critical value of the stellar 
rotation period should increase for smaller planetary masses and larger stellar 
relaxation factors.

\begin{figure}
\begin{center}
\resizebox{1\columnwidth}{!}{\includegraphics{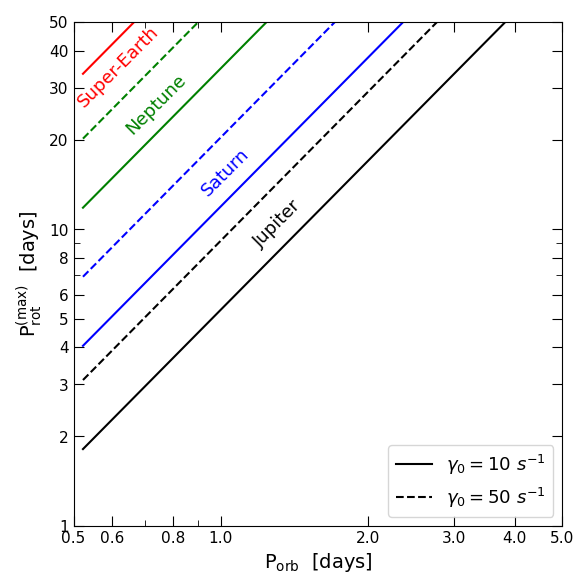}}
\caption{Critical value of the stellar rotation for which magnetic 
braking and tidal spin-up are equal (see equation (\ref{eq6})). Values are 
shown as function of the orbital period of the planet. Different colored lines 
correspond to different planet types. Continuous lines show result assuming 
$\gamma_0 = 10$ $s^{-1}$, while for dashed lines we used $\gamma_0 = 50$ 
$s^{-1}$.}
\label{fig4}
\end{center}
\end{figure}

Figure \ref{fig4} shows the critical value of $P_{\rm rot}$ as a function of 
the orbital period, for four different planet types; each is identified by the 
text alongside the color line. Masses were taken from Table \ref{tab:data} 
while the star was assumed Solar-type with $\alpha_0 = 0.07$ and $k_{f_0} = 
0.05$. Continuous lines correspond to $\gamma_0 = 10$ $s^{-1}$, while their 
dashed counterparts were obtained assuming $\gamma_0 = 50$ $s^{-1}$. 

Jupiter-type planets, characterized by large masses, generate strong tidal 
interactions with the star, thus leading to relatively small values of 
$P^{(\rm max)}_{\rm rot}$ even for large relaxation factors. This is consistent 
with the existence of the moraine-like distribution of hot Jupiters in the P-P 
diagram, especially noticeable for orbital periods below $P_{\rm orb} \sim 3$ 
days. 

\begin{figure*}
\begin{center}
\resizebox{2\columnwidth}{!}{\includegraphics{
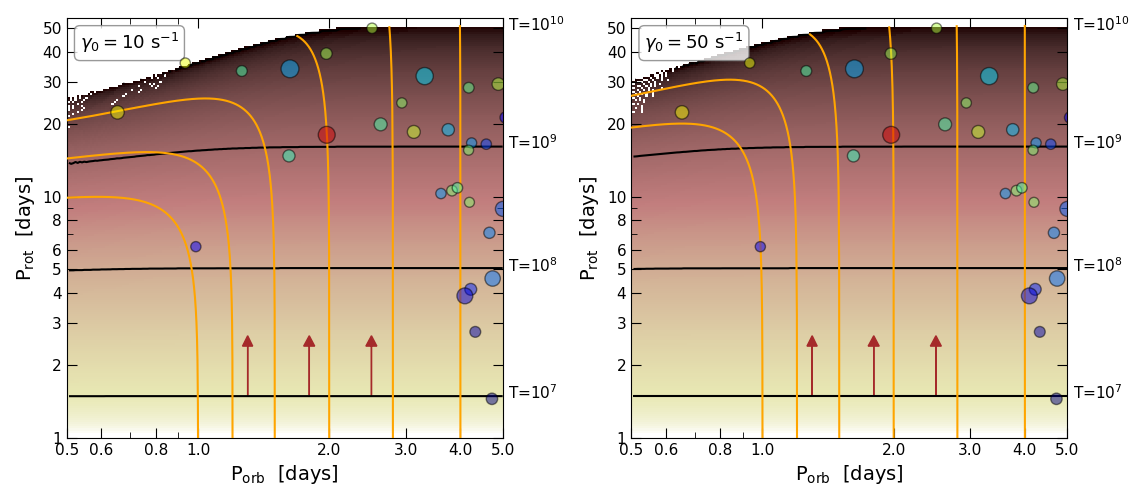}}
\caption{Same as Figure \ref{fig3}, but considering planets with radius $R 
\in [3,10] R_\oplus$. Since most of these planets orbit cool stars, especially 
for $P_{\rm orb} \lesssim 3$, the tidal evolution was simulated assuming $m_0 = 
0.8 m_\odot$, $R_0 = 0.77 R_\odot$ and $\alpha_0 = k_{f_0} = 0.05$ 
\citep{Claret.2019}.}
\label{fig5}
\end{center}
\end{figure*}

\begin{figure*}
\begin{center}
\resizebox{2\columnwidth}{!}{\includegraphics{
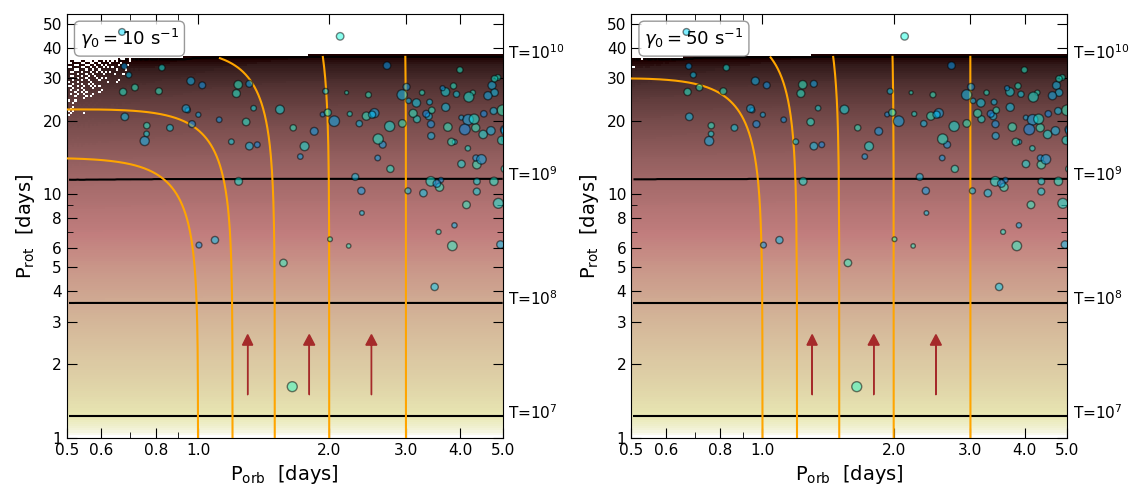}}
\caption{Same as Figure \ref{fig4}, but now focusing on small-size planets 
$R < 3 R_\oplus$ around Solar-type stars. Adopted values for the stellar 
relaxation factor $\gamma_0$ are indicated on the top left-hand corner of each 
plot.}
\label{fig6}
\end{center}
\end{figure*}

The case of smaller gaseous/icy planets (Saturn and Neptune) is less clear. 
Although the results of Figure \ref{fig4} seem to predict a moraine-type 
distribution, at least for Saturn-size bodies and for very small orbital 
periods, the observability also depends on the observed distribution of 
planets. This is analyzed in Figure \ref{fig5} for two different values of 
$\gamma_0$. As was done in Figure \ref{fig3} for hot Jupiters, the orange 
curves show the evolutionary curves of six initial conditions with different 
$P_{\rm orb}$ and the same stellar rotation period $P_{\rm rot} = 1$ day. 
Orbits are assumed circular and with $n = \Omega_1$. 

The distribution of confirmed planets and Kepler candidates is again shown 
with filled circles whose colors are indicative of the effective temperature of 
the star. The paucity of bodies in this size range with orbital period less 
than 2-3 days is well known \citep{Beauge.Nesvorny.2013} and usually referred 
to as the sub-Jovian desert. It is interesting to note that most of the 
candidates at this orbital distance belong to cool stars with $T_{\rm eff} 
\lesssim 5500 \;K$. Consequently, in our simulations of tidal+braking evolutions 
we assumed a central star with $m_0 = 0.8 m_\odot$, $R_0 = 0.77 R_\odot$ and 
$\alpha_0 = k_{f_0} = 0.05$ \citep{Claret.2019}.

Given the lack of sub-Jovian planets very close to the star, it is difficult to 
correlate their distribution with our dynamical model, let alone discuss which 
tidal relaxation factor better fits the data. However, except for a single 
system with $P_{\rm orb} \simeq 1$ day and orbiting a fast rotator, the rest of 
the population does not show any inconsistency with the expected evolutionary 
tracks, nor does it seem necessary to include additional phenomena into the 
model. 

Finally, Figure \ref{fig6} shows analogous results, but now focusing on small 
planets ($R \le 3 R_\oplus$). Since the known population is large, we 
restricted the analysis to Solar-type stars. We verified that there is no 
significant difference with the results obtained from bodies around cooler or 
hotter stars. Again we assume that the close-in planets reached this region 
when the star was still a fast rotator. As with the hot Jupiters described 
previously, the main scenarios for the origin of these systems is planetary 
migration and planet-scattering following disk dispersal 
\citep[e.g.][]{Izidoro.etal.2021}. Both processes are believed to have occurred 
before magnetic braking had the opportunity to slow the stellar rotation 
significantly.

As expected from the values of $P^{(\rm max)}_{\rm rot}$ predicted 
from Figure \ref{fig4}, no moraine-type structure is observed; moreover the 
distribution in the P-P diagram appears almost flat for timescales between 
$10^9$ and $10^{10}$ years. 

As discussed recently by \cite{Messias.etal.2022}, we also find no relevant 
evidence of the lower bound for orbital periods, at least in the form suggested 
by \cite{McQuillan.etal.2013}. The rapid slow-down of $P_{\rm rot}$ due to 
magnetic braking and the tidal decay for very short-period planets seems to 
account for the observed distribution quite well. Both values of $\gamma_0$ lead 
to similar outcomes, although the shape of the tracer orange curves seems like a 
better fit for $\gamma_0 = 10$ $s^{-1}$. For orbital periods $P_{\rm orb} 
\gtrsim 3$ days, tidal effects are negligible and the rotation of the 
corresponding stars evolve following only the rotational braking due to stellar 
winds. The time scale in this case follows closely the Skumanich law 
\citep{Skumanich.1972}.

\section{The star's relaxation factor}\label{sec:Q}

In our simulations for Jupiter--size planets we considered three different 
values for the stellar relaxation factor $\gamma_0$: $2$, $10$ and $50$ $s^{-1}$ 
(see Figure \ref{fig3}). They show that, at least for Solar-type stars, a value 
of$\gamma_0$ between $\sim 10$ $s^{-1}$ and $\sim 50$ $s^{-1}$ appears to 
better represents the possible evolution of these systems to reach the 
moraine-like domain. Smaller values lead to evolutionary tracks in which the hot 
Jupiter falls on the star before the system reaches the moraine-like 
accumulation.
 
This result may be compared to those obtained by \cite{Ferraz-Mello.etal.2015}
for the rotation of the host stars in several systems with large companions 
(exoplanets or brown dwarfs). There, the result was rather centered on 
$\gamma_0 = 50$ $s^{-1}$ because of the adoption of a value of the fluid Love 
number $k_{f_0}$ some 5 times larger than the value adopted here.

While the distribution of smaller planets did not allow to further constrain 
$\gamma_0$, the observed structure are consistent with the values derived from 
hot Jupiters. Similarly, the distribution of sub-Jovian planets around cooler 
stars do not appear to show any significant difference with respect to that 
expected for Solar-type stars. We thus found no evidence that the relaxation 
factor of stars could be a strong function of the stellar type.

\begin{figure}
\begin{center}
\resizebox{0.9\columnwidth}{!}{\includegraphics{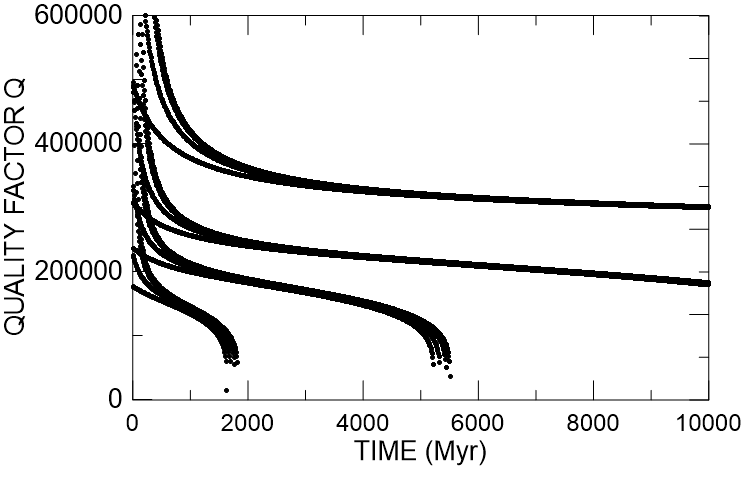}}
\caption{Variation of the stellar quality factor $Q_0$ along the evolutionary 
paths of fig. \ref{fig3}, calculated with $\gamma_0 = 10$ $s^{-1}$. Only the 
parts of the paths in which $\Omega_0 < n$ are shown. Because of the adopted 
value $k_{f_0}=0.05$, the often used alternative quantity $Q'_0$ is 30 times 
larger than $Q_0$.} 
\label{fig7}
\end{center}
\end{figure}

It is interesting to see how the above results are translated in terms of the 
stellar quality factor $Q_0$ (or its variant $Q'_0=1.5 Q_0/k_{f_0}$) used in 
some current tidal friction theories. Using the relation given in section 
\ref{sec:Qdef} and supposing $|\nu| \ll \gamma_0$, we may write
\be
Q_0 = \frac{\gamma}{|\nu|} = \frac{\gamma_0}{2|n-\Omega_0|} .
\label{eq8}
\ee
we obtain the curves shown in fig. \ref{fig7}. We note the big variation of 
$Q_0$ along the path in all solutions making evident that the choice of $Q_0$ 
to parametrize the dissipation in evolving systems is not a good one. The 
definition of $Q_0$ mixes a property of the body, the relaxation, with two 
frequencies of the system, $\Omega_0$ and $n$. Finally, we omitted in 
fig. \ref{fig7} the parts of the paths in which $\Omega_0 < n$ and the 
neighborhood of the synchronization where the definition of $Q_0$ becomes 
singular. 

\section{Conclusion}

This paper shows that the distribution of the points representing the 
short-period Kepler systems and KOIs ($P < 5$ days), for which the period of 
rotation of the star and the orbital period could be determined, presents some 
features that may be explained by the joint action of the tidal evolution of the 
system and the magnetic braking of the star. The main features in the two 
periods (P-P) diagram are: (1) The existence of a moraine-like accumulation of 
systems hosting a hot Jupiter with orbital period in the range $1.5 - 5$ days, 
all in an inclined zone around the stellar rotation period 25 days; (2) The 
absence of any correlation between orbital period and stellar rotation in the 
case of small ($R \lesssim 3 R_\oplus$) planets. 

The creep tide theory allowed us to calculate the evolution of systems with one 
exoplanet in orbit around a Sun-like star showing that systems hosting a hot 
Jupiter with orbital period shorter than 5 days have a fast evolution upwards in 
the P-P diagram and stops evolving exactly in the moraine-like accumulation seen 
in the diagram or, if the initial orbital period is much shorter, fall on the 
star. In the case of systems hosting small planets, tidal evolution quickly 
pulls down planets with orbital periods initially smaller than 1.5-2 days; 
those with periods slightly larger first evolve upwards thanks to the magnetic 
braking and then slowly fall towards the star. The distribution of these systems 
in the P-P diagram is consistent with a flow ruled by tidal evolution and 
magnetic braking. 

The boundaries of distributions seen in the P-P diagram are determined by the 
intensities of these two agents. In the case of the tidal evolution, we have 
found that the most probable relaxation factor of the Sun-like star lies between 
$10$ $s^{-1}$ and $50$ $s^{-1}$. If the dissipation is much larger or much 
smaller than these values, the accumulation would not be located in the place 
where they are observed in the P-P diagram constructed with either the results 
of \cite{McQuillan.etal.2013} or the results of \cite{Santos.etal.2019} and 
\cite{Santos.etal.2021}. This result is in agreement with the results obtained 
for the stars relaxation factor by \cite{Ferraz-Mello.etal.2015}, if we take 
into account that the fluid Love number used there ($0.26$) was some $5$ times 
larger than the value used in this paper ($0.05$) and that, for gaseous bodies, 
the variation of the orbital elements is proportional to $k_f/\gamma$. In this 
paper, we used a determination of the Sun's $k_f$ obtained with the density 
profile of the standard solar model and the formulas given by 
\cite{Folonier.etal.2015} to compute the fluid Love number of layered 
non-homogeneous bodies. 

The use of so-called equivalent initial semimajor axis $a_{\rm equiv}$ in our 
simulations allowed us to avoid the lack of information regarding the primordial 
eccentricities of the planets as they arrived to the region close to the star. 
Finally, the lack of a moraine-like structure or any significant correlation 
between $P_{\rm rot}$ and $P_{\rm orb}$ for low-mass planets is also in 
accordance with our model, and seems additional and strong evidence that 
tidal interactions, together with magnetic braking, have probably been the 
driving forces behind many of the observed dynamical features of close-in 
planets and their host stars.

\subsection*{Acknowledgements}
The authors wish to thank the insightful analysis of an anonymous referee that
helped improve this work, both in content and presentation. SFM acknowledges the 
support of CNPq (Proc. 303540/2020-6) and FAPESP (Proc. 2016/13750-6 ref. 
Mission PLATO). CB is grateful to SECYT/UNC for financial support.

\subsection*{Data Availability}
All data analyzed in this paper is publicly available, but a copy may be 
requested from the authors. 

\bibliography{MCQ}{}
\bibliographystyle{mnras}

\appendix
\section{The Equations of Motion}\label{sec:eqs}

\subsection{Tidal variation of the elements}
The equations used to calculate the tidal variations of the orbital elements: 
semi-major axis and eccentricity, consider the density profile of the bodies, 
but imposes some constraints in order to avoid a huge number of free parameters.
They assume that the body is formed by co-rotating homogeneous layers
\citep{Ferraz-Mello.etal.2022}. They are 
\begin{equation}
\langle\dot{a}\rangle = -\frac{k_f n  R^2\overline{\epsilon}_\rho}{15a} 
\sum_{k\in \mathbb{Z}} \left( 3(k-2) E_{2,k}^2\sin{2\sigma_{k}}
+ k E_{0,k}^2 \sin{2\sigma''_{k}}\right)
\end{equation}
and
\begin{equation}
\begin{split}
\langle \dot{e} \rangle &=  -\frac{k_fnR^2\overline{\epsilon}_\rho}{30a^2 e } 
(1-e^2) \\ 
&\times \sum_{k\in \mathbb{Z}} 
\Big[3\Big(\frac{2}{\sqrt{1-e^2}}+(k-2)\Big)E_{2,k}^2\sin{2\sigma_{k}} 
+ k E_{0,k}^2 \sin{2\sigma''_{k}} \Big] ,\\
\end{split}
\label{eq:dot_e}
\end{equation}
where $\overline{\epsilon}_\rho$ is the mean flattening of the equivalent Jeans 
homogeneous spheroid: 
\begin{equation}
\overline{\epsilon}_\rho = \frac{15MR^3}{4ma^3}. \ \ \ \ \ \ \ \ \ \ 
\label{eq:ep}
\end{equation}
In this expression $m$ is the mass of the tidally deformed body, $M$ is the 
mass of the companion whose attraction is creating the tidal potential, $R$ is 
the radius of the deformed body  and $a$ is the semi-major axis of the relative 
orbit of the two bodies. No hypothesis is done on the relative size of the two 
masses. In fact, in the general case, we have to consider the tides raised in 
the star by the planet and also the tides raised in the planet by the star, so 
$m \ll M$ and $M \ll m$. In both cases we use the same equations and we have 
added their contributions to the variations of the orbital elements to get the 
total effect. 

In the applications described in this paper, both were considered, but the 
contribution of the tides on the exoplanet are too small to be significant. The 
role of the tides on the exoplanet is to quickly drive the rotation of the 
planet to the stationary almost synchronous state where it remains trapped for 
the rest of the time (see Fig. \ref{fig2}).

The tidal equations involve some known functions:
\begin{itemize}
\item The Cayley functions of the orbital eccentricity $e$: 	
\begin{equation}
E_{q,p}(e) = 
\frac{1}{2\pi}\int_0^{2\pi}\left(\frac{a}{r}\right)^3\cos{(qv+(p-q)\ell)\ 
d\ell};
\label{eq:cayley}
\end{equation}
where $r$ is the modulus of the position vector, $\ell$ is the mean anomaly and 
$v$ is the true anomaly. For small eccentricities, we may use the polynomial 
approximations published by \cite{Cayley.1861} or, equivalently, the Hansen 
coefficients \citep{Hughes.1981}. 

\item
The other functions are
\begin{equation}
\begin{split}
\sin{2\sigma_{k}} &= \frac{2\gamma(\nu+kn)}{{\gamma^2+(\nu+kn)^2}}; \\   
\sin{2\sigma''_{k}} &= \frac{2\gamma kn}{{\gamma^2+k^2 n^2}}; \\
\end{split}
\end{equation}
where $\gamma$ is the relaxation factor and $\nu=2\Omega - 2n$ is the mean 
semi-diurnal frequency ($\Omega$ is the rotational velocity of the deformed 
body and $n$ is the orbital mean motion).
\end{itemize}

The other parameter appearing in the variation equations is the fluid Love 
number,  $k_f$. This parameter is related to the mass concentration of the body 
and may be determined from a model of the density profile of the body 
\citep{Folonier.etal.2015}. It is equal to twice the apsidal motion constant 
introduced in the study of close binary stars \citep[see][]{Kopal.1953}. 
In the case of the present-day Sun, the standard solar model leads to a value 
close to 0.05. In previous applications, a rule using the relationship between 
the fluid Love number and the moment of inertia ($k_f=15 C/ 4mR^2$) was used; 
however, this relation, valid for  homogeneous bodies, overestimates the 
value of $k_f$ in the cases under study in this paper.

The variation of the angular velocity of rotation of the deformed body is given 
by 
\begin{equation}
\langle \dot{\Omega} \rangle = - \frac{GMmR^2k_f\overline{\epsilon}_\rho}{5Ca^3}
\sum_{k\in \mathbb{Z}} E_{2,k}^2 \sin{2\sigma_{k}} ,
\end{equation}
where $C$ is the moment of inertia of the deformed body and $G$ the gravitation 
constant.

\subsection{Tidal Dissipation}\label{sec:Qdef}
 The tidal energy released inside the star is negligible when compared to the 
thermal  energy produced by the hydrogen burning near the center of the star. 
However, many authors use the quality factor $Q$ to parameterize the tidal 
effects and the equation giving the tidal dissipation is necessary to allow the 
comparison of the relaxation factor of the creep tide theory to the quality 
factor used in several current tidal friction theories. 
 
In the creep tide theory, the variation of the mechanical energy is
\begin{equation}
\begin{split}
\langle\dot{W}\rangle &= -\frac{GMm R^2 k_f \overline{\epsilon}_\rho}{30a^3} 
\times  \\
& \sum_{k\in \mathbb{Z}} \Big[ \big( 6  (\Omega-n) + 3kn  \big)  
E_{2,k}^2\sin{2\sigma_{k}} + kn E_{0,k}^2 \sin{2\sigma''_{k}}\Big].	\\
\end{split}
\end{equation}
or, in a first approximation,
\begin{equation}\label{eq:Wdot}
	\langle\dot{W}\rangle = 
	-\frac{GMm R^2 k_f \overline{\epsilon}_\rho}{5a^3}
	 (\Omega-n) \sin{2\sigma_{0}} + {\cal O}(e^2)	
 \end{equation}
(N.B. $E_{2,0}\simeq 1$) \citep[see][]{Ferraz-Mello.etal.2022}. This equation is 
the same used in other theories for motions far from the synchronisation when 
we introduce the quality factor $Q$ through 
\be
Q=\frac{2}{\sin 2\sigma_0}= \frac{\gamma}{|\nu|} + \frac{|\nu|}{\gamma} ,
\ee
\label{eq:Qdef}
as discussed, for example, in \cite{Efroimsky.Lainey.2007}.

\subsection{Braking of the star rotation due to stellar activity}

The magnetic braking of the star rotation was computed using the results of 
\cite{Bouvier.etal.1997} for stars with masses in the range $0.5 M_\odot < 
M_s < 1.1 M_\odot $:
\begin{equation}
\dot\Omega_s =	\left\{
\begin{array}{ll}
-B_W\Omega_s^3&{\rm when}\quad \Omega_s\le\omega_{\rm sat}\vspace{3mm}\\
-B_W\omega_{\rm sat}^2\Omega_s\hspace{1cm}& {\rm when}\quad 
\Omega_s>\omega_{\rm sat}
\label{eq:Bouvier}
\end{array}
\right.
\end{equation}
where $\omega_{\rm sat}$ is the value at which the angular momentum loss 
saturates (fixed at $\omega_{\rm sat}  = 3, 8, 14 \, \Omega_\odot$ for 0.5, 
0.8, and 1.0 $M_\odot$ stars, respectively) and $B_W$ is a factor depending on 
the star moment of inertia, mass and radius through the relation
\begin{equation}
	B_W=2.7\times 10^{47} \frac{1}{C_s}\sqrt{\Big(\frac{R_s}{R_\odot} 
\frac{M_\odot}{M_s} \Big)}\qquad \quad   ({\rm cgs \enspace units}),
	\label{eq:Bw}
\end{equation}
\citep{Bouvier.etal.1997}. The subscript $s$ is used to stress the fact that 
the star parameters are being considered. 

The above form of the law is valid after the star has completed its contraction 
(the stellar moment of inertia $C_s$ no longer changes significantly) and is 
fully decoupled from its primeval disk. F-type stars are not expected to be 
affected by the magnetic braking. 

\end{document}